\documentstyle[prl,aps,multicol,psfig,amssymb]{revtex}
\begin{document}
\title{Recurrent swelling of horizontally shaken granular material}
\author{Dirk~E.~Rosenkranz and Thorsten~P\"oschel}
\address{Humboldt-Universit\"at zu Berlin, 
Institut f\"ur Physik, \\
Invalidenstra\ss e 110,
D-10115 Berlin, Germany}
\draft
\date{\today}
\maketitle
\begin{abstract}  
  This Letter reports an experimental study of the dynamics of
  granular materials when vibrated horizontally. Convective motion is
  observed over wide range of parameters, similarly to the known
  effect observed under vertical vibration.  However, under horizontal
  vibration we observe a striking novel effect on the oscillating
  height of the surface of the material: the frequency of the surface
  oscillations are totally decoupled from the frequency of the drive.
  We explain the effect as resulting from an interplay between
  Reynolds dilatancy due to convective motion and mechanical stability
  of the material.
\end{abstract}
\begin{multicols}{2}

When a rectangular container with granular material is shaken either
vertically or horizontally one observes convection. For the case of
vertical shaking this effect has been intensively studied
experimentally, analytically and by means of computer simulations
(s.~e.g.~\cite{vertikalEX,vertikalANA,vertikalMD}). Many results on
vertical shaking are known and there is a certain degree of
understanding of these phenomena. For horizontal shaking convection
cells have been reported only recently~\cite{ESP,PR} and presently
one is far from understanding these complex phenomena.

In the present letter we report a new effect experimentally observed
in horizontally shaken granular material. As far as we know this is
the first dynamical effect where the frequency of forcing decouples
from the frequency of the effect, i.e. the system reveals an inherent
time scale~\cite{ticking}.

The new effect, which we want to focus here, is recurrent swelling of
horizontally shaken material: For a certain range of amplitude $A$ and
frequency $f$ of the sinusoidal motion of the container
$x(t)=A\cos(2\pi\,f\,t)$ and for a certain region of filling level we
observe a time dependent variation of the volume of the material. We
will show that the fluctuations of the height of the material, the
intensity of the convection flow and the energy dissipation rate in
the flowing material are closely related effects.

There are several experimental investigations of exciting
time-dependent phenomena in vertically vibrated
containers~\cite{timedependent,periodedoubling}. While the time scales
of these effects are comparable with the period of forcing, for
horizontal vibration we find a main difference: The effect reported in
this letter has a period which is up to several hundred times larger
than the period of oscillation $f^{-1}$.  Hence we do {\em not}
observe a period doubling scenario as e.g. in~\cite{periodedoubling}.

\begin{figure}[tbp]
\begin{minipage}{8cm}
\centerline{\psfig{file=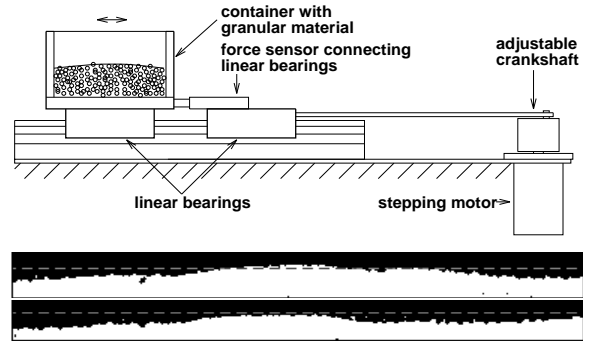}}
  \caption{Top: Experimental setup (see text for explanation). 
    Bottom: Snapshots of the surface region of the material in the
    container. The dashed line leads the eye to a fixed height above
    the container bottom.}
  \label{fig:snap.force}
\end{minipage}
\end{figure}

Fig.~\ref{fig:snap.force} shows the experimental setup which we have
used. The probe carrier was mounted on a precisely balanced horizontal
linear bearing (left in Fig.~\ref{fig:snap.force}). A second bearing
(middle) was driven by a stepping motor via a crankshaft with
adjustable eccentricity. Both bearings were connected by a
piezoelectric force sensor which allows to measure the driving force
with $2~kHz$ time resolution. The motor was computer controlled,
i.e.~in precisely 2,000 impulses the motor axis revolves once. This
high angular resolution provides quasi steady motion. We checked, that
the finite step size per computer signal does not influence the
convective behavior. The entire mechanical device was fixed on an
oscillation damping table.  The container of size $6~\mbox{cm}\times
10~\mbox{cm}$, consists of transparent plastic material. It was
illuminated from the top and was monitored by a videocamera, which was
located in a distance of $4~m$ in a direction perpendicular to the
axis of oscillation. With this camera we observed the dynamical
processes in the container. For that purpose the camera was levelled
exactly at the vertical height of the granular material at rest.

The lower part of Fig.~\ref{fig:snap.force} displays snapshots of the
surface region of width $7~cm$. The container is horizontally shaken
with amplitude $A=0.15~\mbox{cm}$ (the same used in the investigations
described below) and frequency $f=25~\mbox{sec}^{-1}$. As discussed
in~\cite{PR}, we observe a small heap in the center. The heap-height
fluctuates between about $2.3$ and $3.7~mm$ with respect to the level
at rest. The dashed line is drawn $3~mm$ above the level at rest. The
time dependence of the height of the surface on forcing parameters is
subject of our interest. As shown below, the height fluctuates
periodically with a period of about $2.2~\mbox{sec}$, which is 55
times the period of the driving oscillation.

\begin{figure}[htbp]
\begin{minipage}{8cm}
\centerline{\psfig{file=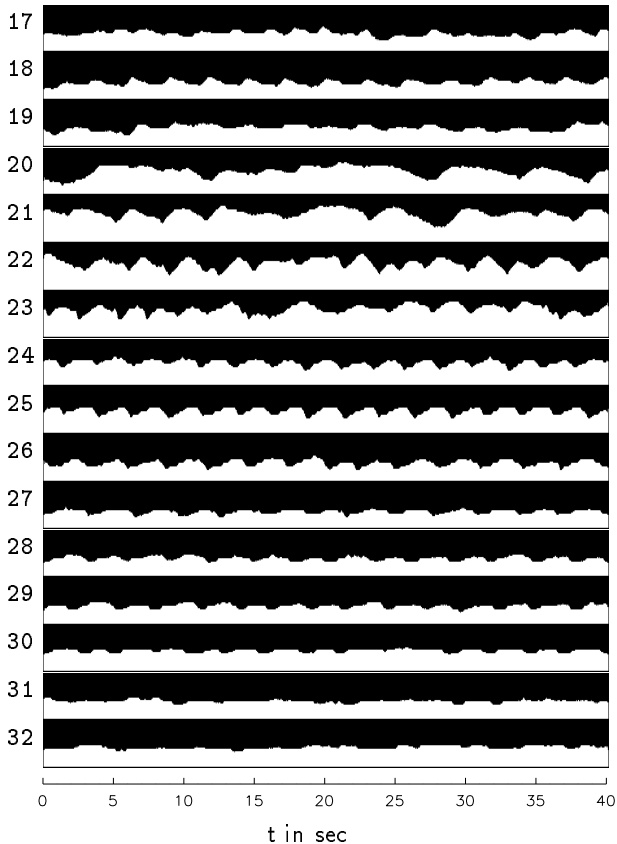}}
\vspace{0.3cm}
  \caption{Height of the granular material in the middle of the 
    container over time for driving frequencies $f$ between
    $17~\mbox{sec}^{-1}$ and $32~\mbox{sec}^{-1}$.}
  \label{fig:assemble.height}
\end{minipage}
\end{figure}

The swelling oscillation of the material has been studied
qualitatively for various combinations of amplitude $A$, frequency $f$
and filling height and for a number of different materials and grain
sizes ($20\mu\mbox{m}\le r\le 100\mu\mbox{m}$). To analyze the effect
we recorded time series of the height of the material surface in the
middle of the container.  Fig.~\ref{fig:assemble.height} displays the
height over a period of $40~\mbox{sec}$ for different driving
frequencies.

\begin{figure}[htbp]
\begin{minipage}{8cm}
\centerline{\psfig{file=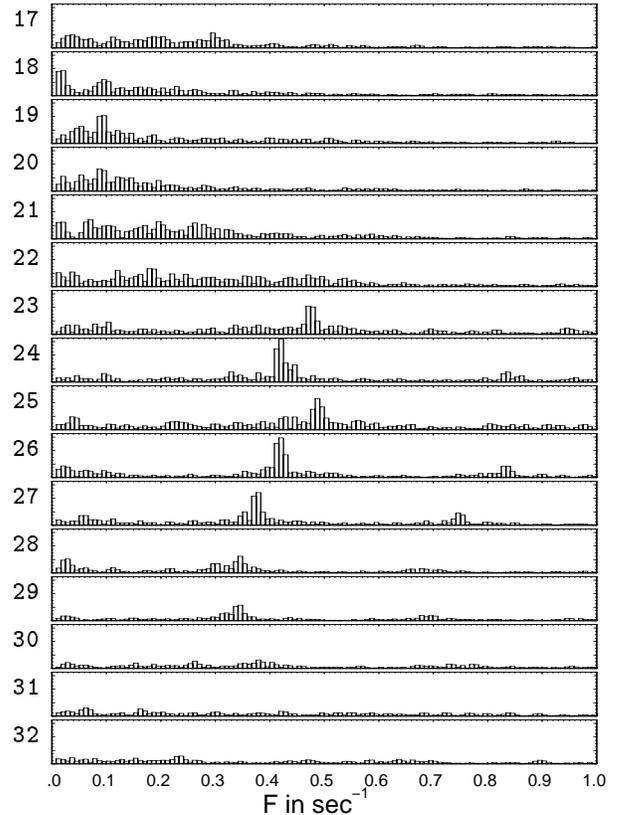}}
\vspace{0.3cm}
  \caption{Fourier transforms of the oscillation of the material 
    height for different driving frequencies corresponding to
    Fig.~2.}
  \label{fig:assemble}
\end{minipage}
\end{figure}

The height of the displayed regions is $0.4~\mbox{cm}$ each. The
material filling height (at rest) was $2.5~\mbox{cm}$, and the average
grain size was $100\,\mu m$. For all frequencies in
Fig.~\ref{fig:assemble.height} one sees that the material height
varies with time. For small driving frequency ($f\lesssim
22~\mbox{sec}^{-1}$) and large driving frequency ($f\gtrsim
30~\mbox{sec}^{-1}$) the height changes irregularly with time, whereas
for $23~\mbox{sec}^{-1} \lesssim f \lesssim 29~\mbox{sec}^{-1}$ we
observe regular, almost periodic oscillation of the material height.
This behavior can be characterized by the Fourier transforms of the
full series of total lengths $120~\mbox{sec}$
(Fig.~\ref{fig:assemble}). For $f < 23~\mbox{sec}^{-1}$ no
characteristic frequency can be observed, i.e. the Fourier spectrum
contains many frequencies.

At $f = 23~\mbox{sec}^{-1}$ there seems to be a sharp transition into
another regime, where we find a characteristic frequency of the height
oscillation as indicated by a peak in the Fourier spectrum. For high
frequency $f \gtrsim 29~\mbox{sec}^{-1}$ the amplitude of swelling
becomes very small and the effect vanishes. Hysteresis of the
transition point has not been observed.

As seen from Fig.~\ref{fig:assemble}, in the region of periodic
swelling ($23~\mbox{sec}^{-1} \lesssim f \lesssim 29~\mbox{sec}^{-1}$)
the swelling frequency $F$ is a function of the driving frequency $f$.

\begin{figure}[tbhp]
\begin{minipage}{8cm}
\centerline{\psfig{file=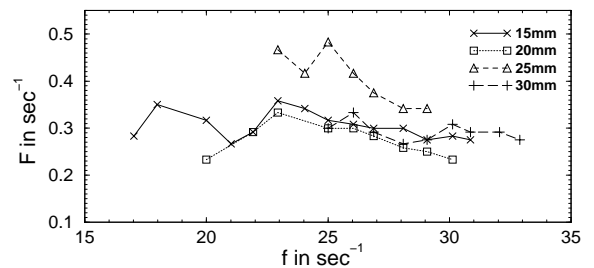}}
        \caption{Swelling frequency $F$ as a function of 
          the frequency of forcing for various filling heights.}
  \label{fig:FoverF}
\end{minipage}
\end{figure}

Fig.~\ref{fig:FoverF} shows the characteristic swelling frequency $F$
over the driving frequency $f$ for different filling heights. The data
shown in Fig.~\ref{fig:FoverF} can be reproduced with good accuracy,
although it seems not to be possible to collapse the data using
dimensionless numbers for many combinations of the system parameters.

In the shaken container one observes convection: the material flows
downwards close to the walls and upwards in the center of the
container. In the interval of driving parameters considered here,
there is an intensive convective material flow. Viewing the container
from top one observes that the oscillation of material height
(Figs.~\ref{fig:assemble.height},\ref{fig:assemble}) corresponds to a
varying particle velocity at the upper surface. When the material is
swelt there is an intensive flow, whereas the flow in the center of
the surface comes almost to rest when the material is collapsed. A
simple way to qualitatively visualize this effect is to take a series
of snapshots of the container viewing from top using a digital camera.

\begin{figure}[htbp]
\begin{minipage}{8cm}
\centerline{\psfig{file=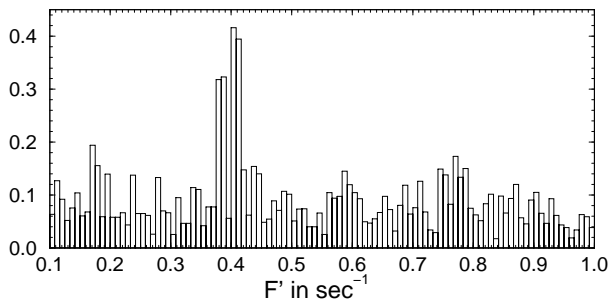}}
  \caption{Material flow at the surface (central area of size 
    $6.5\,cm\,\times 2\,cm$) over time for $f=25~\mbox{sec}^{-1}$.}
\label{fig:ViewFromTop}
\end{minipage}
\end{figure}

In this way one produces a series of difference pictures by
subtracting the gray scale values of the pixels of consecutive
snapshots. The average value $\Delta G$ of the gray scale of the
pixels of the difference pictures provides a measure for the material
flow at the surface.  Fig.~\ref{fig:ViewFromTop} shows the Fourier
transform of $\Delta G$ for a driving frequency of
$f=25~\mbox{sec}^{-1}$. We find that the variation of the surface flow
has the same characteristic frequency as the height oscillation
(c.f.~Fig.~\ref{fig:assemble}). For this reason we believe that
convective motion in the horizontally shaken container and swelling
are closely related effects.

\begin{figure}[htbp]
\begin{minipage}{8cm}
\centerline{\psfig{file=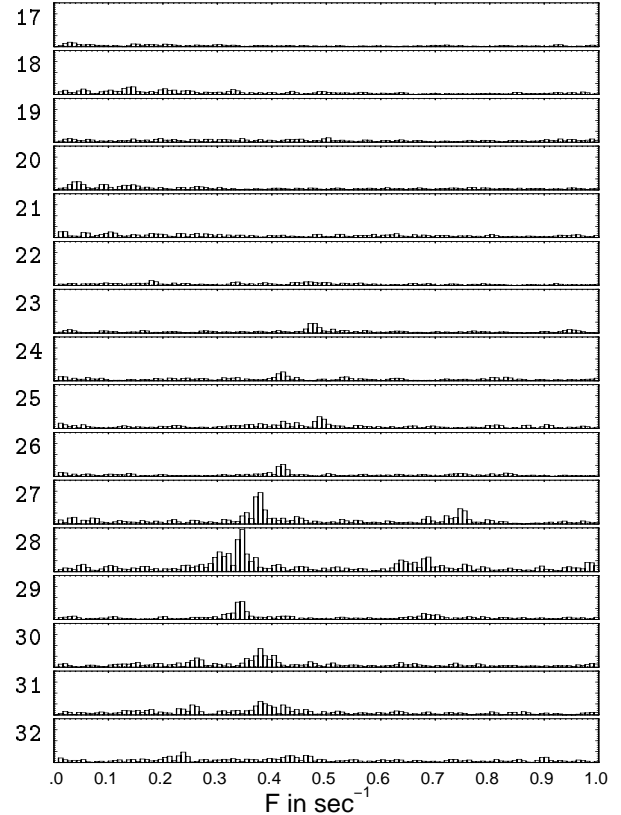}}
\vspace{0.3cm}
  \caption{Fourier spectra of the dissipated energy $E$ for a 
    range of frequencies.}
\label{fig:E}
\end{minipage}
\end{figure}

From the force data $F(t)$ measured by the sensor we calculated the
dissipated driving energy during the $n$th shaking period:
\begin{equation}
  \label{eq:diss}
  E_n=2\pi f A \,\int\limits_{n/f}^{(n+1)/f} 
  F(t)\,\cos\left( 2\pi f t\right)\,dt\,.
\end{equation}
$E_n$ is the energy dissipated during the $n$th period by the entire
system, i.e. by the linear bearing and by the dissipative interaction
of the grains in the container. The time series $E_n$ can be analyzed
by Fourier analysis similar to the surface flow intensity and the
height of the material. Figure~\ref{fig:E} shows the Fourier transform
of the dissipated energy, again for amplitude $A=0.15~\mbox{cm}$ and a
range of driving frequencies $f$.

Comparing Figs.~\ref{fig:assemble} and~\ref{fig:E} we notice, that
both figures agree well, i.e. the characteristic frequencies of height
oscillation and of energy dissipation coincide. In conjunction with
the other experimental observations this leads to the assumption, that
the swelling-effect is essentially connected with a time-dependent
convection, thus a time-dependent material flow and energy dissipation
inside the granular material.

The influence of air has to be discussed. One could imagine that the
convective motion of sand ``pumps'' air into the bulk of the material
which leads to swelling. When the amount of air reaches a certain
extend a ``bubble'' escapes and the material collapses. This mechanism
would explain the swelling effect as well as the saw tooth shape of
the height over time curves (Fig.~\ref{fig:assemble.height}). But
reducing the pressure to $p=50~\mbox{Pa}$, which corresponds to a mean
free path of $\lambda_{air}\approx 130~\mu\mbox{m}$, the swelling
effect {\em does not disappear}. Since $\lambda_{air}\approx
130~\mu\mbox{m}$ is smaller than the typical free path of sand grains
we can exclude air as playing the major role in recurrent swelling.

The origin of the new effect is not completely clear yet, we suggest a
tentative explanation based on the measurement of dissipated
mechanical energy and surface flow: In the horizontally shaken
container one observes convection~\cite{ESP}, i.e. motion of the
particles with respect to each other. To allow for macroscopic motion
in a granular material the material has to be diluted below a certain
density $\rho_R$ before (Reynolds dilatancy~\cite{Reynolds}). When the
local density of the material is below $\rho_R$ the material can start
to flow. The convection in the container implies shear flow, which
causes further dilution and fluidization
(e.g.~\cite{SpahnSchwarzKurths:1997}). When the shaken material swells
according to Reynolds dilatancy, the material becomes diluted and
becomes at the same time less stable mechanically. At a certain moment
the material becomes diluted to an extend that it loses mechanical
stability and collapses. Then the material starts again to swell.

Therefore, we believe that there are two competing effects. First, due
to shear the material tends to dilute. Second, the material has to
remain mechanically stable, i.e. the grains on top have to be
supported by the grains below them and these have to be stabilized by
the grains of the next layer etc.

Our explanation is supported by the measurements of the dissipation of
mechanical energy in the system and of the surface material flow. If
the material is collapsed, i.e. if a large part of the granular
material has a density above $\rho_R$, the grains in these regions
cannot move with respect to each other due to the Reynolds dilatancy
effect.  Hence, in these regions there is no shear motion and,
therefore, none or just low dissipation of mechanical energy. From
this consideration we conclude, that the collapsed material should
dissipate less energy per time than the swelled material. This
behavior coincides with what we found in the measurement of the
dissipated energy. Moreover our results show, that the intensity of
the surface flow oscillates with the same swelling frequency $F$.
Therefore, the experimental results on energy dissipation and surface
flow support our hypothesis on the origin of the swelling effect.

Other interesting questions concern the dependence of the swelling
effect on container shape and size as well as on the grain material
and the amplitude of shaking. We did experiments with several types of
containers, with different grain material and with various amplitudes.
While the properties of swelling depend strongly on the details of the
experiment we observed the effect for a wide range of parameters. A
more detailed description of the properties of the swelling effect
will be subject of a forthcoming paper.

The authors thank R.~P.~Behringer, E.~Clement, J.~A.~Freund,
J.~A.~C.~Gallas, H.~J.~Herrmann and H.~M.~Jaeger for helpful
discussion and V.~Buchholtz, R.~Lieske, K.~Reinhardt and M.~Stock for
help with mechanic equipment. The work was supported by Deutsche
Forschungsgemeinschaft (grants DFG Po~472/2 and DFG Ro~548/5).

\end{multicols}
\end{document}